\documentclass[10pt, aps, prl, nofootinbib, 
               amsmath, amssymb, twocolumn,
               preprintnumbers, showpacs,
               raggedbottom,
               floatfix]{revtex4-1}

\newcommand{\QMC}{\textsc{qmc}}
\newcommand{\DFT}{\textsc{dft}}

\newcommand{\SLDA}{\textsc{slda}}
\newcommand{\ETF}{\textsc{etf}}
\newcommand{\UNEDF}{\textsc{unedf}}
\newcommand{\SciDAC}{\textsc{s}ci\textsc{dac}}
\newcommand{\DoE}{\textsc{d}o\textsc{e}}
\newcommand{\LDRD}{\textsc{ldrd}}
\newcommand{\LANL}{\textsc{lanl}}
\newcommand{\NERSC}{\textsc{nersc}}
\newcommand{\MMFGRANT}{\textsc{de-fg02-00er41132}}
\newcommand{\SGGRANTa}{\textsc{de-fc02-07er41457}}
\newcommand{\SGGRANTb}{\textsc{de-ac52-06na25396}}
\newcommand{\AGGRANTa}{\textsc{de-fg02-97er41014}}
\newcommand{\AGGRANTb}{\textsc{de-ac52-06na25396}}

\providecommand{\abs}[1]{\lvert{#1}\rvert}
\providecommand{\norm}[1]{\lVert#1\rVert}
\DeclareMathOperator{\sech}{sech}

\usepackage[breaklinks]{hyperref}
\usepackage{graphicx}
\graphicspath{{./figures/euler/}}
\usepackage{dcolumn}
\usepackage{calc}

\usepackage{lettrine}

\usepackage{microtype}
\usepackage{xcolor}
\usepackage{bm}

\usepackage[sc,osf]{mathpazo}
\usepackage[euler-digits,small]{eulervm}
\AtBeginDocument{\renewcommand{\hbar}{\hslash}}

\LettrineOptionsFor{T}{lines=3,
        lhang=0.9,
        findent=0.5em,
        nindent=-0.1\LettrineWidth-0.5em}

\usepackage{booktabs}

\usepackage{siunitx}
\renewcommand{\figurename}{Figure}
\renewcommand{\tablename}{Table}
\makeatletter
\renewcommand{\fnum@figure}{\textbf{\figurename~\thefigure}}
\renewcommand{\fnum@table}{\textbf{\tablename~\thetable}}
\makeatother

\begin{document}

\title{Resonantly Interacting Fermions In a Box}

\author{Michael McNeil Forbes,$^{1,2,3}$ Stefano Gandolfi,$^3$ and Alexandros
  Gezerlis$^{2,3}$}

\affiliation{$^1$Institute for Nuclear Theory, University of Washington,
  Seattle, Washington 98195--1560 USA}
\affiliation{$^2$Department of Physics, University of Washington, Seattle,
  Washington 98195--1560 USA}
\affiliation{$^3$Theoretical Division, Los Alamos National Laboratory, Los
  Alamos, New Mexico 87545, USA}

\date {\today}

\begin{abstract}
  \noindent
  We use two fundamental theoretical frameworks to study the finite-size (shell)
  properties of the unitary gas in a periodic box: 1) an \textit{ab initio}
  Quantum Monte Carlo (\QMC) calculation for boxes containing 4 to
  130 particles provides a precise and complete characterization of the
  finite-size behavior, and 2) a new Density Functional Theory (\DFT) fully
  encapsulates these effects.  The \DFT\ predicts vanishing shell structure for
  systems comprising more than 50 particles, and allows us to extrapolate the
  \QMC\ results to the thermodynamic limit, providing the tightest bound to date
  on the ground-state energy of the unitary gas: $\xi_S \leq 0.383(1)$.  We also
  apply the new functional to few-particle harmonically trapped systems,
  comparing with previous calculations.
\end{abstract}
\preprint{\textsc{la-ur 10-07272}}
\preprint{\textsc{int-pub-10-060}}
\preprint{\textsc{nt@uw-10-24}}
\maketitle
\lettrine{T}{he fermion many-body problem} plays a fundamental role in a vast
array of physical systems, from dilute gases of cold atoms to nuclear physics in
nuclei and neutron stars.  The universal character of this problem -- each
system is governed by a similar microscopic theory -- coupled with direct
experimental access in cold atoms, has led to an explosion of recent interest
(see Refs.~\cite{IKS:2008, *giorgini-2007} for a review).  Despite this broad
applicability, we are far from fully understanding even the simplest system: the
``unitary gas'' comprising equal numbers of two fermionic species interacting
with a resonant $s$-wave interaction of infinite scattering length $a_{s} \to
\infty$.  Lacking any scale beyond the total density $n_{+} = n_a + n_b$, the
unitary gas eschews perturbative expansion and requires experimental measurement
or accurate numerical simulation for a quantitative description -- the latter is
presently more precise.  Typical Quantum Monte Carlo (\QMC) calculations,
however, can access at most a few hundred particles.  Density Functional Theory
(\DFT) provides a complementary approach through which one may extrapolate these
results to large systems beyond the reach of direct simulation.

In this Letter, we present the most precise \QMC\ calculations to date of the
unitary gas in a periodic box, studying from 4 to 130 particles, thereby
providing a benchmark for many-body theories.  We use this to calibrate a local
\DFT, then use this \DFT\ to study the finite-size effects (``shell'' effects in
nuclear physics) and extrapolate the \QMC\ results to the thermodynamic limit.
We provide the most precise bound to date of the universal Bertsch
parameter~\cite{Bertsch:1999:mbx, *Baker:1999:PhysRevC.60.054311,
  *baker00:_mbx_chall_compet} $\xi_S = \mathcal{E}/\mathcal{E}_{FG} \leq
0.383(1)$.  ($\mathcal{E}_{FG} = 3/5 n_{+} E_{F}$ is the energy density of a
free Fermi gas with the same total density $n_{+} = k_F^3/(3\pi^2)$, and $E_{F}
= \hbar^2 k_F^2 / (2m)$ is the Fermi energy.)  We also explore the finite-size
properties of the \DFT\ -- a crucial element in the program to calculate
properties of finite nuclei with a universal \DFT~\cite{UNEDF}.  We find that a
local \DFT\ can capture the finite-size effects in these systems without the
need for particle number projection.  We limit our discussion to symmetric
systems ($n_a = n_b$), leaving odd-even staggering to future work, as the \DFT\
then requires an additional dimensionless parameter to characterize the
asymmetry $n_a \neq n_b$.

The \QMC\ results presented here are directly applicable to cold $^6$Li or
$^{40}$K atoms, and constrain dilute neutron matter in neutron
stars~\cite{Gezerlis;Carlson:2008-03}; likewise, the general \DFT\ approach has
myriads of applications throughout cold-atom and nuclear physics (see
Ref.~\cite{Bulgac:2011} for a review).  Our calculation of $\xi_S$ is consistent
with previous results, but an order of magnitude more precise.  Continuum \QMC\
bounds $\xi_S \lesssim 0.40$ -- $0.44$ with an uncertainty no better than the
last digit~\cite{CCPS:2003, ABCG:2004, Carlson:2005kg, Gezerlis;Carlson:2008-03,
  Gezerlis:2009, Morris:2010}.  Lattice \QMC\ results range from $\xi_S \approx
0.3$ -- $0.4$~\cite{BDM:2008, Lee:2008, *Abe:2009, Carlson:private_comm2,
  *Drut:2010, *Endres:2010}, comparable to analytic
results~\cite{Haussmann:2007,*Nishida:2007}.  Experimental groups found
qualitative agreement~\cite{Bartenstein;Altmeyer;Riedl;Jochim;Chin...:2004-05,
  *KTTCSL:2005, *PLKLH:2005}, which led to precision measurements: notably with
Duke~\cite{Luo:2009} and Paris~\cite{Navon:2010} quoting $0.39(2)$ and $0.41(1)$
respectively.

D\textsc{ft} is an in principle exact approach, widely used in in nuclear
physics~\cite{Drut:2010kx,*Gezerlis:2010,*Gandolfi:2010}, and in quantum
chemistry to describe normal (i.e., non-superfluid) systems. It has recently
been extended to describe the unitary gas~\cite{Papenbrock:2005fk, Bulgac:2007a,
  Rupak:2008fk, *Salasnich:2008, Bulgac:2011}. We build upon one approach -- the
Superfluid Local Density Approximation (\SLDA) -- which was originally
constrained by \QMC\ calculations of the continuum state, and then validated
with \QMC\ calculations in a harmonic trap~\cite{Chang;Bertsch:2007-08,
  Blume;Stecher;Greene:2007-12} (see also Fig.~\ref{fig:trap}).  We focus on
translationally invariant systems in a periodic box to isolate the finite-size
effects from the gradient effects.  We find that the inclusion of an anomalous
density is crucial: functionals attempting to model the superfluid by adding
only gradient or kinetic corrections~\cite{Papenbrock:2005fk, Rupak:2008fk,
  *Salasnich:2008} are unable to even qualitatively characterize the finite-size
effects.

Our \QMC\ results are based on a fixed-node Diffusion Monte Carlo approach that
projects out the state of lowest energy from the space of all wave functions
with fixed nodal structure as defined by an initial many-body wave function
(ansatz).  By varying the ansatz, we obtain a variational upper bound on the
ground-state energy.
\begin{figure}[t]
  \centering
  \includegraphics[width=\columnwidth]{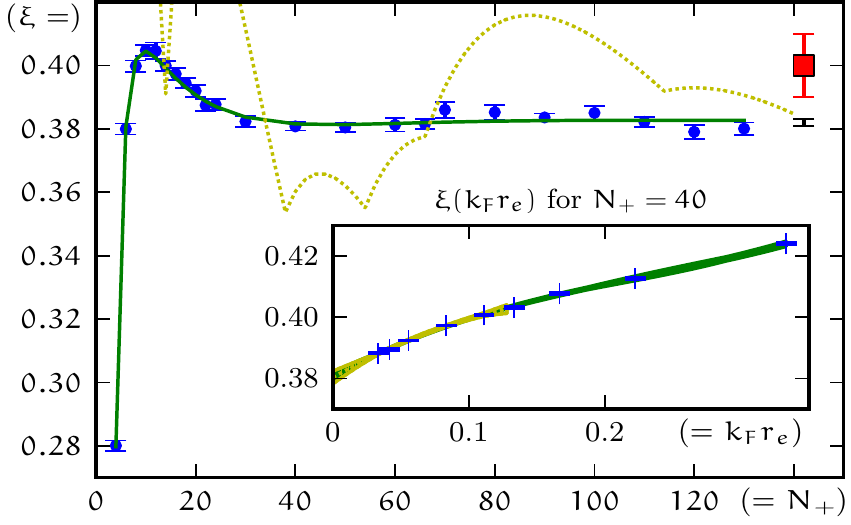}
  \caption{(color online) Ground-state energy-density $\xi =
    \mathcal{E}/\mathcal{E}_{FG}$ of $N_+$ fermions in a periodic cubic box at
    the unitary limit.  The circles with error bars are the result of using a
    quadratic least-squares extrapolation to zero effective range of our new
    \QMC\ results.  The solid curve is the best fit \SLDA\ \DFT.  The light
    dotted curve is the functional considered in~\cite{Papenbrock:2005fk} with
    $\alpha=0.69$. For comparison, we have plotted the previous best estimate
    $\xi_S=0.40(1)$ (red square) and the current estimate $\xi_S=0.383(1)$ below
    it to the far right of the figure. Inset: we show the typical
    effective-range dependence $\xi(k_Fr_e)$ with the best fit 1$\sigma$ error
    bounds for all-point cubic (solid dark green) and five-point quadratic
    (hatched light yellow) polynomial fits.  Note that: a) the five-point
    quadratic model is consistent with the full cubic model and has a comparable
    extrapolation error, and b) the inflection point near $k_Fr_e \approx 0.16$
    necessitates a higher-order fit for larger ranges (cubic is sufficient for
    the ranges shown here). Results for $N_+=40$ show the same qualitative
    behaviour; hence, for the other points we use the five-point quadratic
    extrapolation.}
  \label{fig:ener}
\end{figure}%
In this work, we use the trial function introduced in~\cite{CCPS:2003}:
\begin{equation}
  \Psi_T = 
  \mathcal{A}[
  \phi(\mathbf{r}_{11'})\phi(\mathbf{r}_{22'})\cdots\phi(\mathbf{r}_{nn'})]
  \prod_{ij'}f(r_{ij'}),
\end{equation}
where $\mathcal{A}$ antisymmetrizes over particles of the same spin (either
primed or unprimed) and $f(r)$ is a nodeless Jastrow function introduced to
reduce the statistical error. The antisymmetrized product of $s$-wave pairing
functions $\phi(\mathbf{r}_{ij'})$ defines the nodal structure:
\begin{equation}
  \phi(\mathbf{r}) = 
  \sum_{\mathbf{n}}\alpha_{\norm{\mathbf{n}}}e^{ i {\mathbf{k}}_{\mathbf{n}} \cdot
    \mathbf{r}}
  +
  \tilde{\beta} (r).
  \label{eq:bcsphideco}
\end{equation}
The sum is truncated (we include ten coefficients) and the omitted short-range
tail is modelled by the phenomenological function $\tilde{\beta}(r)$ chosen to
ensure smooth behavior near zero separation.  We use the same form for
$\tilde{\beta}(r)$ as in~\cite{CCPS:2003} with the values $b=0.5$ and $c=5$, and
vary the 10 coefficients $\alpha_{\norm{\mathbf{n}}}$ for each $N_+$ to minimize
the energy as described in Ref.~\cite{Sorella:2001}.  Representative nodal
structures are defined by the coefficients in Table \ref{tab:wf}.  The same
ansatz suffices for different effective ranges, but an independent optimization
is required for each $N_+$.

\begin{table}[t]
  \footnotesize
  \begin{tabular}{
      S
      S[table-text-alignment=right]
      S[table-text-alignment=right,color=gray]
      S[table-text-alignment=right,
        round-mode=places,
        round-precision=4,
        color=gray]}
    \toprule
    {\normalsize $N_+$} & 
    {\normalsize $\xi(N_+)$} &
    {\normalsize $\xi_{\text{box}}(N_+)$} &
    {\normalsize $\xi_{N}/\xi_{\text{box}}$}\\
    \midrule
      4 & 0.280 +- 0.002 & 0.205 +- 0.001 & 1.36414\\
      6 & 0.380 +- 0.002 & 0.274 +- 0.001 & 1.38804\\
      8 & 0.400 +- 0.002 & 0.310 +- 0.001 & 1.28903\\
     10 & 0.405 +- 0.002 & 0.342 +- 0.001 & 1.18491\\
     12 & 0.405 +- 0.003 & 0.370 +- 0.002 & 1.09301\\
     14 & 0.400 +- 0.002 & 0.394 +- 0.002 & 1.01445\\
     16 & 0.397 +- 0.002 & 0.367 +- 0.002 & 1.08272\\
     18 & 0.394 +- 0.002 & 0.355 +- 0.001 & 1.11217\\
     20 & 0.392 +- 0.002 & 0.350 +- 0.002 & 1.11967\\
     22 & 0.387 +- 0.002 & 0.348 +- 0.002 & 1.11442\\
     24 & 0.388 +- 0.002 & 0.352 +- 0.002 & 1.10169\\
     30 & 0.382 +- 0.002 & 0.366 +- 0.002 & 1.04435\\
     40 & 0.381 +- 0.001 & 0.393 +- 0.001 & 0.96985\\
     50 & 0.380 +- 0.001 & 0.391 +- 0.001 & 0.97256\\
     60 & 0.381 +- 0.002 & 0.386 +- 0.002 & 0.98685\\
     66 & 0.382 +- 0.002 & 0.383 +- 0.002 & 0.99498\\
     70 & 0.386 +- 0.003 & 0.379 +- 0.003 & 1.01768\\
     80 & 0.385 +- 0.002 & 0.368 +- 0.002 & 1.04605\\
     90 & 0.384 +- 0.001 & 0.365 +- 0.001 & 1.04978\\
    100 & 0.385 +- 0.002 & 0.370 +- 0.002 & 1.04027\\
    110 & 0.382 +- 0.002 & 0.373 +- 0.002 & 1.02359\\
    120 & 0.379 +- 0.002 & 0.373 +- 0.002 & 1.01729\\
    130 & 0.380 +- 0.002 & 0.375 +- 0.002 & 1.01388\\
    \bottomrule
  \end{tabular}
  \caption{Values of $\xi(N_+)=E(N_+)/V/\mathcal{E}_{FG}$ shown in
    Fig.~\ref{fig:ener}.  To facilitate comparison with other normalizations in
    the literature, we include the values $\xi_{\text{box}}(N_+)=
    E(N_+)/E_{FG}(N_+)$ where $E_{FG}(N_+)$ is the energy of $N_+$
    non-interacting particles in a box.  The conversion factor is shown in the
    last column.}
  \label{tab:results}
\end{table}

\begin{table}[b]
  \newcommand{\colwidth}{2.1em}
  \newcolumntype{.}{D{.}{.}{2}}
  \newcolumntype{w}{p{\colwidth}}
  \newcommand{\aI}[1]{\multicolumn{1}{c}{\normalsize{$a_{#1}$}}}
  \renewcommand{\t}[1]{#1}
  \begin{center}
    \setlength{\tabcolsep}{3.612pt}
    \begin{tabular*}{\columnwidth}{rcccccccccc
      }
      \toprule
      {\normalsize $N_+$} & \aI{0} & \aI{1} & \aI{2} & \aI{3} & \aI{4} & 
      \aI{5} & \aI{6} & \aI{8} & \aI{9} & \aI{10}\\[2pt]
      \midrule
      10 & 1600 & 350 & 49 & 16 & 12 & 14 & 14 & 11 & 9.0 & 6.7\\[2pt]
      40 & 160 & 91 & 27 & 0.49 & -2.8 & -0.086 & 2.2 & 2.9 & 2.5 & 1.9\\[2pt]
      80 & -24 & 13 & 12 & 8.2 & 5.1 & 3.7 & 2.7 & 2.0 & 1.6 & 1.0\\[2pt]
      120 & -51 & -17 & 0.51 & 7.8 & 6.3 & 5.8 & 4.6 & 2.5 & 1.7 & 1.0\\[2pt]
      \bottomrule
    \end{tabular*}
  \end{center}
  \caption{\label{tab:wf}
    Sample coefficients of the pairing function~(\ref{eq:bcsphideco})
    $\alpha_{\norm{\mathbf{n}}} = 10^{-4}a_{I}$ where $I = \norm{\mathbf{n}}^2 =
    n_x^2 + n_y^2 + n_z^2 = k^2L^2/4\pi^2$\kern-0.4em.
    Higher-order coefficients are set to zero.}
\end{table}

We simulate the Hamiltonian:
\begin{equation}
  \mathcal{H} = \frac{\hbar^2}{2m}\Biggl(- \sum\limits_{k = 1}^{N_+} \nabla_k^{2}
  \;-\; 4v_0 \mu^2\sum_{i,j'}\sech^2(\mu r_{ij'})\Biggr),
\end{equation}
with an interspecies interaction of the modified P\"{o}schl-Teller type
(off-resonance intraspecies interactions are neglected).  We tune to infinite
$s$-wave scattering length by setting $v_0=1$: the effective range becomes $r_e
= 2/\mu$.  To extrapolate to the zero-range limit, we simulate at $\mu/k_F \in
\{12.5, 24, 36, 48, 60\}$ for which $0.03 < k_Fr_e < 0.16$.  A careful
examination of additional ranges up to $k_Fr_e \sim 0.35$ for $N_+=40$ and
$N_+=66$ (see the inset in Fig.~\ref{fig:ener}) reveals that a three-parameter
quadratic model in $r_e$ is necessary and sufficient to extrapolate our results
without a systematic bias; the results are shown in Fig.~\ref{fig:ener}.%

The energies exhibit definite finite-size effects for $N_+ \lesssim 50$, but are
essentially featureless for larger $N_+$.  This lack of structure is confirmed
by the best fit \DFT\ (discussed below) and disagrees with the results presented
in Ref.~\cite{Morris:2010}.  The values of $\xi$ for $N_+ > 50$ are distributed
about the best fit value $\xi_S \approx 0.383(1)$, and represent the lowest
variational bounds to date. Part of the decrease from previous results is due to
the careful extrapolation to zero effective range.  The remainder is due to the
improved optimization of the variational wave function.

To model the finite-size effects we turn to a local \DFT\ for the
unitary Fermi gas that generalizes the \SLDA\ originally presented in
Ref.~\cite{Bulgac:2007a}.  In addition to the total density $n_+ =
2\sum_{n}\abs{v_{n}}^2$, the \SLDA\ includes both kinetic
$\tau_+=2\sum_n\abs{\nabla v_n}^2$ and anomalous densities $\nu=\sum_{n}
u_{n}v_{n}^{*}$.  (The $+$ index signifies the sum of the contributions coming
from the two components $a$ and $b$; $u_{n}(\mathbf{r})$ and
$v_{n}(\mathbf{r})$ are the Bogoliubov quasiparticle wave functions.)  The
original three-parameter \SLDA\ is
\begin{equation}
    \label{eq:DF_SLDA}
    \mathcal{E}_{\SLDA} =
    \frac{\hbar^2}{m}\left(
      \frac{\alpha}{2}\tau_{+} +
      \beta \frac{3}{10}(3\pi^2)^{2/3}n_{+}^{5/3}\right) +
    g\nu^{\dagger}\nu,
\end{equation}
where $\alpha$ is the inverse effective mass; $\beta$ is the self-energy; and
$\gamma$ controls the pairing through the regularized coupling
$g=1/(n_{+}^{1/3}/\gamma - \Lambda/\alpha)$ where $\Lambda\rightarrow\infty$ is
a momentum cutoff that we take to infinity (see Ref.~\cite{Bulgac:2011} for
details).  One can use numerically the equations for homogeneous matter in the
thermodynamic to replace the parameters $\beta$ and $\gamma$ with the more
physically relevant quantities $\xi_S$ and $\eta = \Delta/E_F$, where $\Delta$
is the pairing gap.

In principle, the \DFT\ can be expressed in terms of only the density $n_+$ and
its gradients.  References~\cite{Rupak:2008fk, *Salasnich:2008} consider local
formulations of this type (called Extended Thomas-Fermi (\ETF) functionals).
Since gradients vanish in the periodic box, \ETF\ functionals reduce to
$\mathcal{E}_{\ETF}(n_+) \equiv \xi_S \mathcal{E}_{FG}$ and exhibit no
finite-size structure, contrary to the \QMC\ results.
Reference~\cite{Papenbrock:2005fk} adds $\alpha\tau_+$, but without
$\nu^\dagger\nu$ the finite-size effects do not correlate with the \QMC\
behavior (see Fig.~\ref{fig:ener}) and the best fit to our results is also
flat ($\alpha \rightarrow 0$).  Furthermore, such functionals cannot
qualitatively reproduce the quasiparticle dispersion relationship, an attractive
feature of the \SLDA\ (see also Ref.~\cite{Bhattacharyya:2005}).

The best fit three-parameter \SLDA\ functional (\ref{eq:DF_SLDA}) -- $\alpha =
1.26(2)$, $\xi_S = 0.3826(5)$, and $\eta = 0.87(2)$ -- is shown in
Fig.~\ref{fig:ener}.  It fits the 23 \QMC\ points from $N_+ = 4$ to $N_+ = 130$
with a reduced chi squared $\chi^2_{\text{r}} = 0.7$, indicating complete
consistency.  Although remarkable, the fit is not completely satisfactory: 1)
Fitting the exact two-particle energy $\xi_2 = -0.4153\cdots$ raises
$\chi^2_{\text{r}} = 2.0$, and 2) the best fit gap parameter $\eta$ and inverse
effective mass $\alpha$ are inconsistent with the values $\eta = 0.50(5)$ and
$\alpha = 1.09(2)$ obtained from the $N_+=66$ \QMC\ quasiparticle dispersion
relation~\cite{Carlson:2005kg,BF:2008}, and the values $\eta =
0.45(5)$~\cite{Carlson;Reddy:2008-04} and $\eta =
0.44(3)$~\cite{Schirotzek;Shin;Schunck;Ketterle:2008-08} extracted from
experimental data.

\begin{figure}[t]
  \centering
  \includegraphics[width=\columnwidth]{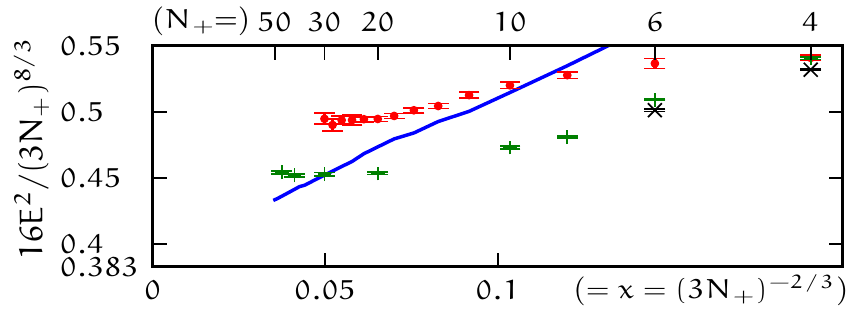}
  \caption{(color online) Ground-state energy of the harmonically trapped
    unitary Fermi gas (in units where $\hbar\omega = 1$) scaled to demonstrate
    the asymptotic form $16E^2/(3N_+)^{8/3} = \xi_S\bigl(1 + cx +
    \mathrm{O}(x^{7/6})\bigr) \smash{{}^2}$ predicted by the low-energy
    effective theory of Ref.~\cite{SW:2006}.  The best fit \SLDA\ (solid blue
    line) is compared with zero-range results for $N_+\in\{4, 6\}$ from
    Ref.~\cite{Blume:2010}, and finite-range \QMC\ results from
    Ref.~\cite{Blume;Stecher;Greene:2007-12} (upper red dots) and
    Ref.~\cite{Stefano_private:2010} (green pluses).  The latter have
    significantly lower energy, despite having a slightly large effective range,
    suggesting that the wave functions in
    Ref.~\cite{Blume;Stecher;Greene:2007-12} were not fully optimized.  We
    expect careful optimization and zero-range extrapolation to bring the \QMC\
    results for large $N_+$ in line with the \DFT\ as discussed in the text.}
  \label{fig:trap} 
\end{figure}

These deficiencies might be remedied by generalizing the \SLDA.  As noted in
Ref.~\cite{Bulgac:2011}, the following combination of divergent kinetic and
anomalous densities is finite:
\begin{equation}
  \label{eq:capK}
  K = \frac{\hbar^2\tau_{+}}{2m} + \frac{g}{\alpha} \nu^{\dagger}\nu
    = \frac{\hbar^2\tau_{+}}{2m} + \frac{\nu^{\dagger}\nu}
                 {\alpha n_{+}^{1/3}/\gamma - \Lambda}.
\end{equation}
The lack of scales thus dictates the functional form:
\begin{align}
  \label{eq:DF_new}
  \mathcal{E}(K, n_{+}) &= \xi(Q)\; \mathcal{E}_{FG}(n_{+}), &
  Q &= K/\mathcal{E}_{FG}(n_{+}),
\end{align}
where $Q$ is dimensionless, and the regularization condition depends on $Q$
through the function $\gamma(Q)$.  The original \SLDA\ is linear $\xi(Q) =
\alpha Q + \beta$ with constant $\gamma(Q)=\gamma$.  This generalized functional
can fit any monotonic $\xi(N_+)$, including the exact $N_+=2$ point.  For $N_+ >
6$, $\xi(N_+)$ is not monotonic and the functional is in principle constrained.
For example, requiring that $\xi=\xi_S$ at both $N_+\approx 6.2(2)$ and
$N_+=\infty$ fixes the ratio $\eta/\alpha = 0.69(2)$.  (As an aside, we note
that the momentum distribution $n_k$ in the \DFT\ relates this to the
``contact'' $C$: $\eta/\alpha = \sqrt{2C}/k_F^2 \approx 0.44$ -- $0.49$; see
Refs.~\cite{Gandolfi:2010a, *Drut:2010a} and references therein, though it is
not clear that this property should be trusted.)  In practice, the errors and
the discreteness in $N_+$ leave room for flexibility in the functional
form, and we have found several generalized functional forms with $\chi_r^2
\approx 1.5$ while constraining $\eta = 0.50$.  We may have to accept the
discrepancy in $\alpha$ as a limitation of the \DFT.

However, generalizing the \SLDA\ may not be needed: analyzing the ``symmetric
heavy-light ansatz''~\cite{Lee:2008a}, (justified by lattice \QMC\
calculations~\cite{Lee:2008,*Abe:2009}), we find that the simple three-parameter
\SLDA\ suffices ($\chi_r^2 \approx 0.5$) with reasonable $\alpha=0.96(2)$,
$\eta=0.51(1)$, and $\xi=0.322(2)$ -- slightly higher than the
$\xi=0.31(1)$ extracted in~\cite{Lee:2008a}.

It is not trivial that the simple \DFT~(\ref{eq:DF_SLDA}) captures all
finite-size effects above $N_+=4$ to high precision in both calculations,
indicating that the \SLDA\ may be used to extrapolate to the thermodynamic
limit.  We note that no particle-number projection is required -- a quite
ill-defined procedure often considered necessary in nuclear
physics~\cite{Duguet:2009}: Perhaps improved nuclear functionals may similarly
capture finite-size effects through local anomalous densities in the spirit of
$\nu$.

To finish, we consider harmonically trapped systems in Fig.~\ref{fig:trap}.  As
discussed in~\cite{SW:2006}, the energy may be expressed as $E(N_+) =
\tfrac{1}{4}\hbar\omega\sqrt{\xi_S}(3N_+)^{4/3} \bigl(1 + cx +
\mathrm{O}(x^{7/6})\bigr)$ where $x = (3N_+)^{-2/3}$ and $c$ is expressed in
terms of low-energy coefficients.  As demonstrated by the zero-range
$N_+\in\{4,6\}$ results of~\cite{Blume:2010}, the \DFT\ still over-estimates the
energy for small systems, most likely because we have omitted the gradient terms
in the functional that vanish in homogeneous systems.

For large $N_+$ the \DFT\ has the expected asymptotic form with intercept $\xi_S
= 0.383$ unlike the finite-range \QMC\ results of
Refs.~\cite{Blume;Stecher;Greene:2007-12, Stefano_private:2010}.  This is
qualitatively consistent with the leading effective-range corrections which
scale asymptotically as $x^{-1/4}$; the systematic overestimation of the energy
by the variational \QMC\ approach might also contribute.  We defer further
discussion until carefully extrapolated zero-range results are published.

To summarize, we present the most precise Quantum Monte Carlo calculations to
date of a symmetric unitary Fermi gas in a periodic box comprising 4 to 130
particles.  By carefully characterizing and extrapolating these results to zero
effective range, we have completely mapped out the finite-size effects.  These
results are used to analyze the structure of a Density Functional Theory for the
symmetric unitary gas, and it is shown that the simplest three-parameter form of
Eq.~(\ref{eq:DF_SLDA}) fully accounts for all shell effects to within the
statistical errors of the \QMC\ results without the need for particle-number
projection; a more complicated form, however, may be required to capture both
the finite-size effects and the quasiparticle dispersions. The \DFT\ predicts no
significant shell corrections beyond 50 particles, and the \QMC\ calculations
confirm this, allowing us to extract a precise upper bound on the universal
equation of state $\xi_S \leq 0.383(1)$, an order of magnitude improvement in
precision over previous bounds and the lowest bound of any variational method to
date.  The functional in its latest form is well constrained, but leads to
slight disagreements with \QMC\ predictions for harmonic traps.  Converging both
\QMC\ and \DFT\ approaches promises to be a fruitful direction of future
research.

\begin{acknowledgments}
  We thank Aurel Bulgac, Joe Carlson, and Dean Lee for useful discussions.  This
  work is supported, in part, by \textsc{us} Department of Energy (\DoE) grants
  \MMFGRANT, \AGGRANTa, \& \AGGRANTb, \DoE\ contracts \SGGRANTa\ (\UNEDF\
  \SciDAC) \& \SGGRANTb, and by the \LDRD\ program at Los Alamos National
  Laboratory (\LANL).  Computations for this work were carried out through Open
  Supercomputing at \LANL, on the \textsc{uw} Athena cluster, and at the
  National Energy Research Scientific Computing Center (\NERSC).
\end{acknowledgments}

%
\end{document}